\documentstyle[preprint,aps,epsf]{revtex}
\begin{document}
\draft
\preprint{\vbox{
\null\hfill KEK preprint 95-179 \\
\null\hfill OCU-HEP 95-02 \\
\null\hfill NWU-HEP 95-03 \\
\null\hfill TUAT-HEP 95-01 \\
\null\hfill DPNU-95-22 \\
\null\hfill PU-95-697 \\
\vskip -4cm\noindent \epsfbox{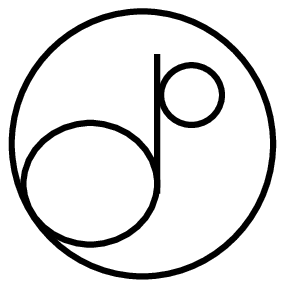}
}}
\title{Observation of Highly Virtual Photon-Photon Collisions
to Hadrons at TRISTAN\footnote{to be published in Phys. Lett. B}
}
\author{\underline{
R.Enomoto$^{(1)}$}\footnote{internet address: enomoto@kekvax.kek.jp.},
K.Abe$^{(2)}$, T.Abe$^{(2)}$, I.Adachi$^{(1)}$,
K.Adachi$^{(3)}$, M.Aoki$^{(4)}$, M.Aoki$^{(2)}$, S.Awa$^{(3)}$,
K.Emi$^{(5)}$, H.Fujii$^{(1)}$, K.Fujii$^{(1)}$,T.Fujii$^{(6)}$,
J.Fujimoto$^{(1)}$,
K.Fujita$^{(7)}$, N.Fujiwara$^{(3)}$, H.Hayashii$^{(3)}$,
B.Howell$^{(8)}$, H.Ikeda$^{(3)}$, R.Itoh$^{(1)}$, Y.Inoue$^{(7)}$,
H.Iwasaki$^{(1)}$,
M.Iwasaki$^{(3)}$, K.Kaneyuki$^{(4)}$, R.Kajikawa$^{(2)}$,
S.Kato$^{(9)}$, S.Kawabata$^{(1)}$, H.Kichimi$^{(1)}$, M.Kobayashi$^{(1)}$,
D.Koltick$^{(8)}$, I.Levine$^{(8)}$, S.Minami$^{(4)}$,
K.Miyabayashi$^{(3)}$, A.Miyamoto$^{(1)}$
K.Nagai$^{(10)}$\footnote{Present address; Dept. of Particle Phys.,
The weizmann Inst. of Sci., Rehovot 76100, Israel},
K.Nakabayashi$^{(2)}$, E.Nakano$^{(7)}$,
O.Nitoh$^{(5)}$, S.Noguchi$^{(3)}$, A.Ochi$^{(4)}$, F.Ochiai$^{(11)}$,
N.Ohishi$^{(2)}$, Y.Ohnishi$^{(2)}$, Y.Ohshima$^{(4)}$,
H.Okuno$^{(9)}$, T.Okusawa$^{(7)}$, E.Shibata$^{(8)}$,
T.Shinohara$^{(5)}$, A.Sugiyama$^{(2)}$,
S.Suzuki$^{(2)}$, S.Suzuki$^{(4)}$, K.Takahashi$^{(5)}$, T.Takahashi$^{(7)}$,
T.Tanimori$^{(4)}$, T.Tauchi$^{(1)}$, Y.Teramoto$^{(7)}$, N.Toomi$^{(3)}$,
T.Tsukamoto$^{(1)}$, O.Tsumura$^{(5)}$, S.Uno$^{(1)}$, T.Watanabe$^{(4)}$,
Y.Watanabe$^{(4)}$, A.Yamaguchi$^{(3)}$, A.Yamamoto$^{(1)}$,
and M.Yamauchi$^{(1)}$\\
(TOPAZ Collaboration)\\
}
\address{
$^{(1)}$KEK, National Laboratory for High Energy Physics,
Ibaraki-ken 305, Japan \\
$^{(2)}$Department of Physics, Nagoya University, Nagoya 464, Japan\\
$^{(3)}$Department of Physics, Nara Women's University, Nara 630, Japan \\
$^{(4)}$Department of Physics, Tokyo Institute of Technology, Tokyo 152,
Japan\\
$^{(5)}$Dept. of Appl. Phys., Tokyo Univ. of Agriculture and
Technology, Tokyo 184, Japan\\
$^{(6)}$Department of Physics, University of Tokyo, Tokyo 113, Japan\\
$^{(7)}$Department of Physics, Osaka City University, Osaka 558, Japan \\
$^{(8)}$Department of Physics, Purdue University,
West Lafayette, IN 47907, USA \\
$^{(9)}$Institute for Nuclear Study, University of Tokyo, Tanashi,
Tokyo 188, Japan \\
$^{(10)}$The Graduate School of Science and Technology,
Kobe University, Kobe 657,
Japan \\
$^{(11)}$Faculty of Liberal Arts, Tezukayama University, Nara 631, Japan \\
}
\date{\today}
\maketitle
\begin{abstract}

We have observed highly virtual ($Q^2>1.05~GeV^2$) photon-photon
collisions to hadronic final states at $\sqrt{s_{e^+e^-}}=58~GeV$.
The integrated luminosity of the data sample was 241pb$^{-1}$.
Both scattered beam-electrons and
scattered beam-positrons were detected using
low-angle calorimeters
(i.e., both photons were highly virtual, ``double-tag");
we obtained 115 hadronic events
with an estimated background of $10.2\pm1.1$.
The cross section obtained was $4.11\pm0.66$pb
in the $2<W_{\gamma \gamma}<25$ GeV and $Q^2_{\gamma, min}>2$ GeV
region, while the lowest order quark-parton model predicted 3.00pb.

\end{abstract}

\narrowtext

A measurement of the total hadronic cross section in highly virtual
photon-photon collisions ($e^+e^-\rightarrow e^+e^- hadrons$) is
reported for center-of-mass energy, $W$, between 2 and 25 GeV.
In this experiment both the scattered $e^+$ and $e^-$, refered as
tags, were detected. Such a technique is called ``double-tag"
\cite{point}.
The range of photon four-momentum squared, $q^2$, for this experiment
was -1.05 to -37 GeV$^2$.
$q^2$'s and $W$ were detected on an event-by-event basis directly from
the tags. Theoretical calculation of such events in which two large
scales of virtuality exist, does not rely on any high $P_T$ selection
in the hadronic final state.
The measurement thus amounts to a first measurement of a well defined
total hadronic cross section in which there could be minimal sensitivity
to QCD cut-offs.

Examples of Feynman diagrams which contribute to $\gamma \gamma$
collisions are shown in Figures \ref{feynman} (a)-(e).
Figure \ref{feynman} (a) is called
multi-peripheral or ``direct process" which is expected
to be dominant in these events \cite{direct}.
Figures 1 (b), (c), and (d) are examples of a bremsstrahlung, conversion,
and annihilation diagram. Figure 1 (e) is an example of
``resolved photon process"  which is considered to be a smaller
contribution in highly virtual photon-photon collisions \cite{resolved}.

The statistics
and virtualities of the photons of the experimental data
are the highest compared with previous experiments \cite{double}.

The integrated luminosity of the event sample is 241pb$^{-1}$ at
$\sqrt{s}=58~GeV$.
The details concerning the TOPAZ detector at TRISTAN
can be found in reference
\cite{topaz}.

To select virtual photon-photon collision events, we
used a pair of low-angle calorimeters (FCL) made of Bismuth Germanate
crystals (BGO) \cite{fcl}. This covered the polar-angle region of 3.2 to
12 degrees with respect to the beam axis. The energy resolution was
measured to be 5\% for Bhabha electrons, and 3\% for 4-GeV electron beams.
We required that both of the FCL's (electron and positron sides) had single
clusters of energies greater than $0.4E_b$ where
$E_b$ is the beam energy.
This cut was determined in order to reduce contributions from
the resolved-photon process \cite{ks} and initial (final) state radiation
in various processes. It also helped us to reject any background from spent
electrons in the accelerator beams.
Under these conditions the virtuality of
the photon ($Q^2=-q^2$) was ensured to be greater than 1.05 GeV$^2$.

The trigger system for the TOPAZ detector consisted of
neutral and charged-track
triggers. The conditions for the neutral triggers were as follows:
(a) The total energy deposited in a barrel calorimeter
(BCL, made of lead-glass)
or an endcap calorimeter (ECL, made of Pb-streamer-tube sandwiches) must be
greater than 2 or 4
(63\% of the events were taken with a 2-GeV threshold of the BCL, and the
others with a 4-GeV threshold.)
and 10 GeV, respectively;
(b) The BCL was segmented into three parts
and the ECL into 4 parts. For energy deposits in these parts, 1 GeV
(for BCL) and 4 GeV (for ECL) thresholds were set. Any two hits caused
trigger signals.
The charged-track trigger required at least two tracks from the origin.
The $P_T$ of each track
was required to be greater than 0.3$\sim$0.7 GeV (run dependent), and the
opening angle between one of the pairs must be greater than 45$\sim$90
degrees. The cut values depended on the beam conditions;
their variation throughout the data taking
were taken into account in Monte-Carlo
simulations. The details can be found in reference \cite{trigger}.

We used a time-projection-chamber (TPC), the BCL,
and the ECL in event selection.
Detailed analyses of these detectors can be found in reference
\cite{nimelectron}.

We required at least three charged tracks from the origin with
a $P_T$ greater than 0.15 GeV.
The invariant mass ($W_{VIS}$) of the visible particles (in the TPC and BCL)
had to be greater than 2 GeV. Here, the energy threshold of the BCL neutral
clusters had to be greater than 100 MeV. The visible energy in this region
had to be less than 25 GeV.

In order to reduce the backgrounds from annihilation events which were
produced at low angle, we required that the energy flow into the ECL
be less than that into the BCL.

The material thickness in front of the TPC was $\sim$0.3 radiation length.
Therefore, we had to pay special attention to $e^+e^-\rightarrow
e^+e^-e^+e^-(\gamma)$.
Using dE/dx measurements from the TPC we identified
electron tracks in the events. To reject this process, we required
that an event must contain at least two non-electron tracks.

Finally, we obtained a total of 115 events.
We investigated three sources of background:
(a) beam-gas and accidental coincidences by spent electrons,
(b) $e^+e^-\rightarrow q\bar{q}(n\gamma)$, and
(c) $e^+e^-\rightarrow e^+e^-l^+l^-(n\gamma)$ where $l$ is a lepton.
\begin{itemize}
\item{(a)} Using the Bhabha-scattering events in the central region of the
detector, as well as random triggered events, we estimated the
background level of the accidental hits in the FCL. The accidental rate
was measured to be less than 0.1\%.
The contamination of the beam-gas scattering was estimated by using
off-vertex events in the experimental sample. It was estimated to be
1.0$\pm$0.7 events.
\item{(b)} The KORALZ generator was used to simulate the initial (ISR) and
final (FSR) state radiation with multiple $\gamma$'s \cite{koralz,miyabayashi}.
Here, the maximum energy ($k_{max}$) of the radiated
photon was set at 0.97$E_b$ for
b-quark events and 0.99$E_b$ for the other quarks.
The next-to-leading order ($\alpha_s$) correction and
hadronization was carried out using JETSET 6.3 \cite{miyabayashi,jetset63}.
The contamination of this process was estimated to be 2.3$\pm$0.7 events.
\item{(c)} The FERMISV generator was used to simulate this process
\cite{fermisv}.
This generator could also simulate ISR and FSR. $k_{max}$ was set at
0.8$E_b$. The contamination was estimated to be 6.9$\pm$0.6 events;
$\tau$-pair events dominantly contributed ($6.0\pm0.5$ events).
\end{itemize}
In total, the background contamination in this sample was
estimated to be $10.2\pm1.1$ events.
We have thus obtained a clean sample of direct photon-photon collision events.

The average of smaller and larger $Q^2$'s of two photons of this
event sample were 5.1 and 12.3 GeV$^2$.
Both are equally highly virtual.
We derived the differential cross sections
of $e^+e^-\rightarrow e^+e^- hadrons$ with respect
to $W_{\gamma \gamma}$ at several $Q^2_{\gamma ,min}$ cuts,
where $Q^2_{\gamma, min}$ was a minimum of two $Q^2_{\gamma}$ detected
by the forward and backward FCL counters.
The cross section such as $\sigma_{\gamma \gamma \rightarrow hadrons}$
or photon structure function would be a future subject in which we
need more statistics and works.
We used the FERMISV code for deriving the acceptance correction
factors with our detector simulation codes.
We again set $k_{max}=0.8E_b$ for ISR and FSR in FERMISV.
We restricted events with all four fermions in the final state having
polar angles greater than 2.5 degrees from the beam axis.
The minimum invariant masses for hadronic system were set to 2 GeV for
light quark events, 2$m_D$ for c, and 2$m_B$ for b, respectively.
The cosines of scattered electrons (positrons)
were required to be within
$0.9762<|\cos \theta|<0.9986$.
Quark masses used were 0.35 GeV for u,d, 0.5 GeV for s, 1.35 GeV for c,
and 4.5 GeV for b.
The hadronization process was simulated using JETSET 6.3 string fragmentation
\cite{jetset63,softpi}.
The obtained cross sections are shown in
Table \ref{tcross} for various kinematic ranges.
Also shown are the ratios between the experimental and theoretical
cross sections and the T-ratios which are the ratio between
$\sigma_{exp}(e^+e^-\rightarrow e^+e^-q\bar{q})
/\sigma_{threory}(e^+e^-\rightarrow e^+e^-\mu^+\mu^-)$,
where only the multi-peripheral diagram is taken into account in calculating
$\sigma_{theory}$.
Radiative corrections were carried out in deriving these values.
The cross section obtained was $4.11\pm0.66$pb
in the $2<W_{\gamma \gamma}<25$ GeV and $Q^2_{\gamma ,min}>2$ GeV
region, while the lowest order (LO)
theory predicted by FEMISV is $3.00$pb.
Figure \ref{fcross} shows the ratios between the experimental and
LO-theoretical cross sections.
Observed distributions such as $W_{VIS}$ and number of charged tracks were
consistent with the Monte-Carlo prediction within statistical errors
except for overall normalizations.
On the other hand, the experimental data agreed with the theoretical
predictions at high $Q^2_{\gamma}$, including overall normalization factors.
The errors shown are both statistical and systematic.
We consider the systematic errors below.

The systematic errors were estimated separately concerning the FCL, trigger,
and event selection.
\begin{itemize}
\item{FCL:} We changed the energy threshold in the clustering algorithm,
as well as that in electron tagging.
The dependence of the acceptance was compared
with a Monte-Carlo simulation. Also, background contamination was
studied by comparing it with Bhabha and random-triggered events.
The systematic error due to the FCL was estimated to be 5.2\%.
\item{Trigger:} We added noise hits in the tracking detector to the
Monte-Carlo simulations. We also studied the acceptance dependence
on the energy thresholds of the neutral triggers. The error was
estimated to be 6.0\% based on the ambiguity in the hardware settings (10\%).
\item{Event selection:}
We changed the cut values and derived
the cross sections.
The differences were taken into account as errors (4.9\%).
\item{Hadronization:}
We used JETSET 6.3  for hadronization of $q\bar{q}$ system \cite{jetset63}.
We changes parameters of string fragmentation $a$ and $\sigma_q$ in the
ranges 0.4-0.7 GeV$^{-2}$ and 0.35-0.45 GeV, respectively. The acceptance
differences were taken into account as systematic errors. Their bin-to-bin
average was 8.9\%.
\end{itemize}
An error of integrated luminosity was estimated to be 4\%.
The errors except for luminosity  and trigger were derived
by a bin-by-bin bases. For example a luminosity error is common for
all bins, i.e., correlates positively. The correlated systematic error was
estimated to be 7.2\%.
The average bin-by-bin systematic errors was estimated to be 13.5\%.
The first errors quoted in Table \ref{tcross} are statistical and the
second ones are systematic. The errors in Figure \ref{fcross} are
quadric sums of these errors.

The accuracy of the FCL energy measurements is important for determining
$W_{\gamma \gamma}$. The FCL energy was originally calibrated at 4 GeV
using AR electron beams at KEK and at 29 GeV by the Bhabha-scattering
electrons. In order to check the other energy region, we studied radiative
Bhabha events. We selected acoplanar Bhabha events where the radiated
$\gamma$ direction was consistent with the FCL fiducial volume.
The agreement between the BCL-predicted and
the FCL-measured energies was as expected from our
Monte-Carlo simulation. The Monte-Carlo simulation
predicted that the $W_{\gamma \gamma}$ resolution would be
almost constant ($\sim$3 GeV) in the
observed energy region ($2<W_{\gamma \gamma}<25$ GeV).
This was better than the $W_{VIS}$
determined from the TPC and BCL.



In conclusion,
we have observed highly virtual ($Q^2>1.05~GeV^2$) photon-photon collisions
to hadronic final states at $\sqrt{s_{e^+e^-}}=58~GeV$.
The integrated luminosity of the data sample was 241pb$^{-1}$.
The beam electrons (positrons) were tagged by low-angle calorimeters;
we obtained $97.3\pm10.3$ events
in the 2 $<W_{\gamma \gamma}<$
25 GeV and $Q^2_{\gamma ,min}>$2 GeV region,
the Monte-Carlo with LO theory predicted $65.6\pm2.2$ events where
the errors are statistical.
In this new kinematic region the contribution of the LO $q\bar{q}$
diagram is expected to dominate.
With systematic errors included,
the ratio between experimental and theoretical data becomes
1.48$\pm$0.26.
The cross sections were measured
to be $4.11\pm0.66$pb
in the region where
$Q^2>2~GeV^2$ and 2 $<W_{\gamma \gamma}
<25~GeV$.

We thank Drs. H. Terazawa, M. Kobayashi,
K. Hagiwara, and T. Izubuchi for helpful discussions.
We also thank the TRISTAN accelerator
staff for the successful operation of TRISTAN.
The authors appreciate
all of the engineers and technicians at KEK as well as
those of the collaborating
institutions: H. Inoue, N. Kimura, K. Shiino, M. Tanaka, K. Tsukada, N. Ujiie,
and H. Yamaoka.

\newpage
\begin{table}
\begin{tabular}{cccccc}
\multicolumn{2}{c}{Kinematic Region} &
\multicolumn{2}{c}{Cross section} & ratio(Exp/Th) & T-ratio\\
$Q^2_{\gamma ,min}$ & $W_{\gamma}$ range
                    & $\sigma_{hadron}^{exp}$
                    & $\sigma_{hadron}^{theory}$
                    & $\sigma_{hadron}^{exp}/\sigma_{hadron}^{theory}$
                    & $\sigma_{hadron}^{exp}/\sigma_{\mu^+\mu^-}^{theory}$ \\
(GeV$^2$) & (GeV) & (pb) & (pb) &  &\\
\hline
 2.0(8.5)& 2.0-6.0(4.6)&2.41$\pm$0.48$\pm$0.33
&1.81&1.33$\pm$0.26$\pm$0.17&1.17$\pm$0.23$\pm$14\\
 2.0(10.7)& 6.0-9.5(7.8)&0.89$\pm$0.20$\pm$0.19
&0.71&1.25$\pm$0.28$\pm$0.26&1.36$\pm$0.30$\pm$0.28\\
 2.0(8.2)& 9.5-13.5(11.4)&0.51$\pm$0.10$\pm$0.09
&0.30&1.69$\pm$0.32$\pm$0.32&1.79$\pm$0.34$\pm$0.36\\
 2.0(8.6)&13.5-25.0(18.6)&0.295$\pm$0.056$\pm$0.060
&0.186&1.59$\pm$0.30$\pm$0.33&1.85$\pm$0.35$\pm$0.39\\
 3.5(9.7)& 2.0-7.0(5.2)&1.03$\pm$0.23$\pm$0.15
&1.00&1.04$\pm$0.23$\pm$0.15&1.01$\pm$0.22$\pm$0.15\\
 3.5(10.5)& 7.0-12.0(9.4)&0.47$\pm$0.10$\pm$0.09
&0.35&1.34$\pm$0.28$\pm$0.25&1.49$\pm$0.31$\pm$0.28\\
 3.5(10.8)&12.0-25.0(16.2)&0.185$\pm$0.039$\pm$0.039
&0.143&1.29$\pm$0.27$\pm$0.28&1.45$\pm$0.30$\pm$0.32\\
 5.0(12.9)& 2.0-9.0(6.4)&0.62$\pm$0.12$\pm$0.15
&0.67&0.93$\pm$0.18$\pm$0.22&0.99$\pm$0.19$\pm$0.23\\
 5.0(12.3)& 9.0-25.0(14.2)&0.179$\pm$0.030$\pm$0.051
&0.184&0.97$\pm$0.16$\pm$0.27&1.12$\pm$0.19$\pm$0.31\\
\end{tabular}
\caption{
Hadronic cross sections
in various kinematic ranges.
The first errors are statistical and the second ones are systematic
errors as described in the text. The terms shown in this table are specified
in the text.
The values in parentheses in the first and second columns are event average
of the kinematic variables.
}
\label{tcross}
\end{table}
\newpage
\narrowtext
\begin{figure}
\caption{Examples of Feynman diagrams which contribute to $\gamma \gamma$
collisions; (a) multi-peripheral, (b) bremsstrahlung, (c) conversion,
(d) annihilation, and (e) ``resolved photon process".
}
\label{feynman}
\vskip 2cm
\caption{
Ratios of experimental and theoretical cross sections
in
various kinematic regions;
(a) $Q^2_{\gamma ,min}>$ 2GeV$^2$,
(b) $Q^2_{\gamma ,min}>$ 3.5GeV$^2$,
and (c) $Q^2_{\gamma ,min}>$ 5GeV$^2$.
The $Q^2_{\gamma , min}$ is specified in the text.
}
\label{fcross}

\end{figure}
\newpage
\section*{Figure 1, Phys. Lett. B., R. Enomoto et al.}
\vskip 3cm
\epsfbox{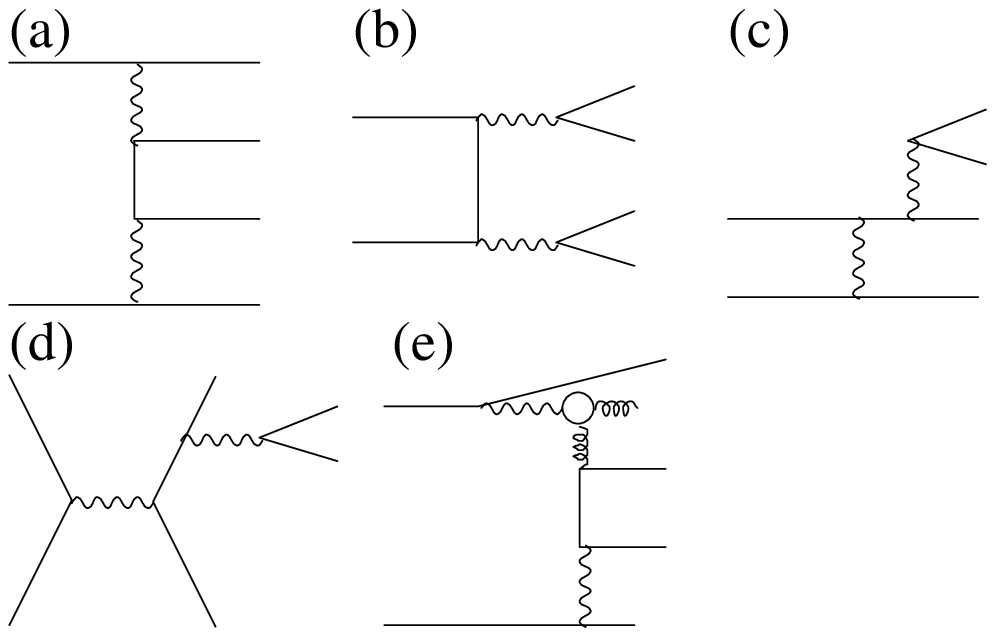}
\newpage
\section*{Figure 2, Phys. Lett. B., R. Enomoto et al.}
\vskip 3cm
\epsfbox{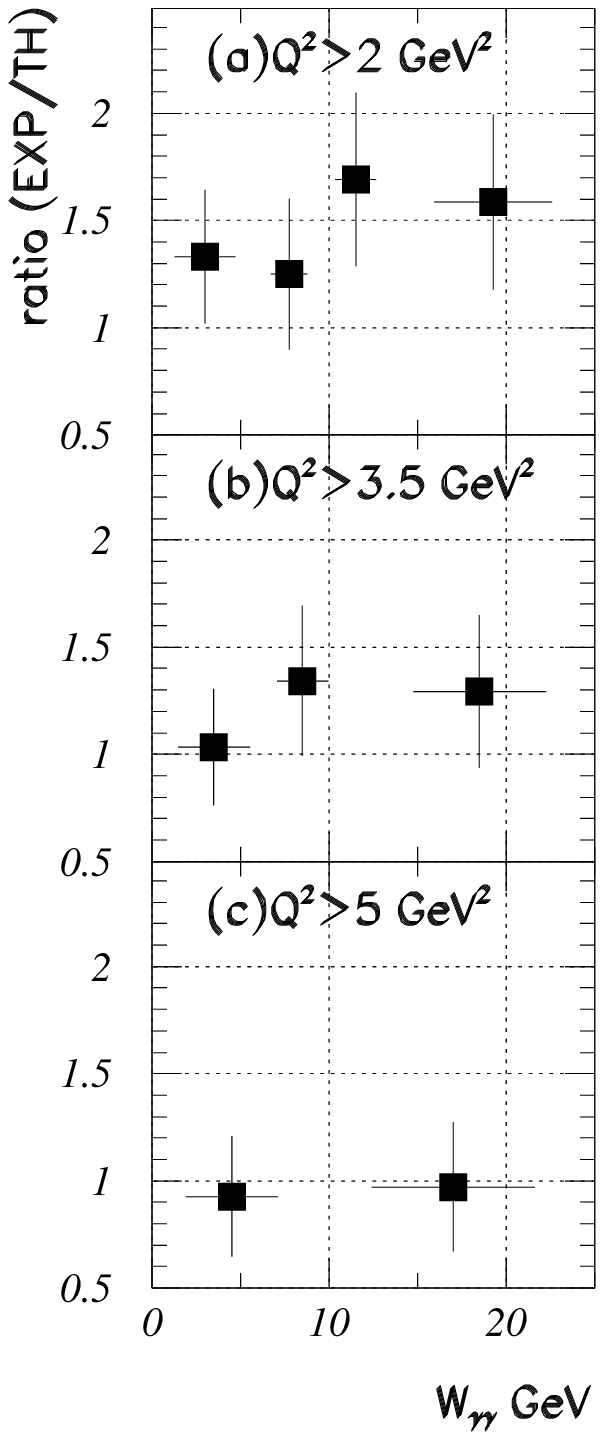}

\end{document}